\documentclass[preprint,twoside,a4paper,11pt]{article}

\usepackage{irrj2e}

\usepackage{caption}
\usepackage{adjustbox} %
\usepackage{tikz,pgfplots}
\usepackage{stfloats}
\usepackage{booktabs}
\usepackage{amsmath}
\usepackage{float}
\usepackage{placeins}

\usepackage{etoolbox}
\usepackage{siunitx}
\usepackage[normalem]{ulem}

\robustify\bfseries
\robustify\uline

\newcommand{\siuline}[1]{\uline{#1}}
\newcommand{\sibf}[1]{\bfseries{#1}}

\usepackage{lastpage}

\usepackage{latexsym}

\usepackage[T1]{fontenc}

\usepackage[utf8]{inputenc}

\usepackage{microtype}

\usepackage{inconsolata}

\usepackage{graphicx}

\usepackage{kotex}
\usepackage{adjustbox} %
\usepackage{tikz,pgfplots}
\usepackage{arydshln}
\usepackage{stfloats}
\usepackage{booktabs}
\usepackage{multirow}
\usepackage{amsmath} 
\usepackage{color,soul}
\pgfplotsset{compat=1.18}
\usetikzlibrary{pgfplots.colormaps}

\usepackage{xcolor}

\irrjheading{1}{2025}{1-\pageref{LastPage}}{2/25}{-}{10.54195/irrj.00000}

\ShortHeadings{The Pre-Training Study of Expanded-SPLADE Models on Web Document Titles}{Kim, Lee and Won}
\firstpageno{1}

\begin{document}

\author{\name Hiun Kim \email hiun.kim@navercorp.com \\
       \addr Naver\\
       Seongnam, South Korea
       \AND
        \name Tae Kwan Lee \email taekwan.lee@navercorp.com \\
       \addr Naver\\
       Seongnam, South Korea
       \AND
        \name Taeryun Won \email lory.tail@navercorp.com \\
       \addr Naver\\
       Seongnam, South Korea}

\editor{My editor}

\title{The Pre-Training Study of Expanded-SPLADE Models on Web Document Titles}

\maketitle

\begin{abstract}

Masked Language Modeling~(MLM) pre-training is one of the primary ways to initialize Neural Information Retrieval (IR) models prior to retrieval fine-tuning.
However, studies show that MLM pre-trained models have limited readiness and transfer learning issues for fine-tuning them into Neural Bi-Encoder models.
This paper studies the effect of different pre-training datasets and pre-training options on the MLM pre-trained models for retrieval fine-tuning.
The study focuses on the SPLADE-style model, which uses the MLM layer also at fine-tuning time.
More specifically, we experimented with Expanded-SPLADE~(ESPLADE) models, a specific instance of SPLADE models, and in-house web document titles are used as datasets.
Pre-training, fine-tuning, and evaluation with optional test-time pruning of sparse vectors are conducted.

Our observations are three-fold:
First, fine-tuned models of higher retrieval effectiveness at both unpruned and most strict pruned settings are mostly pre-trained on a general corpus, and pre-trained with a higher learning rate, showing lower MLM accuracies.
Second, in the most strict pruned setting, those models show higher-level retrieval cost and a higher variance in the length of the individual postings list.
Third, the repetition of the general pre-training dataset does not have much effect on retrieval effectiveness. %
The experimentation empirically identifies the potential limitations for aligning MLM pre-training to ESPLADE fine-tuning. %
Also, the experimentation provides an empirical observation that, at most strict pruned settings, the retrieval effectiveness is better maintained by the higher-level retrieval cost, showing the trade-off relationship between the two in our setting. %

\end{abstract}


\section{Introduction}

Neural Information Retrieval (IR) models develop representation learning for IR tasks~\citep{mitra2018introduction}.
Masked Language Modeling~(MLM) pre-training~\citep{devlin2019bert} can be included as a preparatory component in such models~(e.g.~\citet{lin2022pretrained, formal2021splade}).
However, the MLM pre-training and the task of IR are not always aligned.
\citet{gao2021condenser} studied the limit of the MLM pre-trained BERT for bi-encoder fine-tuning.
They suggest that MLM pre-trained BERT is not pre-trained for aggregate information for bi-encoder fine-tuning.

\citet{lassance2023experimental} shows that MLM pre-training on a retrieval fine-tuning corpus for the IR tasks can result in comparable performance overall to the general pre-trained models.
For example, they show that on SPLADE~\citep{formal2021splade}, a model based on a 6-layer DistilBERT, pre-trained from scratch using the MSMARCO corpus, has overall similar levels of retrieval effectiveness compared to the pre-trained 6-layer DistilBERT pre-trained with a general corpus.

MLM pre-training is, however, still used for some models.
For instance, SPLADE models~\citep{formal2021splade} often use MLM output to make a sparse representation.
This paper studies the effect of different pre-training datasets and pre-training options on the MLM pre-trained models for retrieval fine-tuning.

The study focuses on the SPLADE-style model, which uses the MLM layer also at fine-tuning time.
More specifically, we experimented with Expanded-SPLADE~(ESPLADE) models~\citep{dudek2023learning}, a specific instance of SPLADE models, and in-house web document titles are used as datasets.
A method of \citet{dudek2023learning} is to provide a custom output vocabulary that can have a larger dimension to the original model.
\citet{kim2025role} found that such a property can be particularly useful for the trade-off between retrieval effectiveness and efficiency by test-time pruning of sparse vectors~\citep{lassance2023static, lassance2024two}, thereby, can be practically intriguing.

Pre-training, fine-tuning, and evaluation with optional test-time pruning of sparse vectors are conducted.
Our observations are three-fold:
First, fine-tuned models of higher retrieval effectiveness at both unpruned and most strict pruned settings are mostly pre-trained on a general corpus, and pre-trained with a higher learning rate, showing lower MLM accuracies.
Second, in the most strict pruned setting, those models show higher-level retrieval cost and a higher variance in the length of the individual postings list.
Third, the repetition of the general pre-training dataset does not have much effect on retrieval effectiveness. %

The experimentation empirically identifies the potential limitations for aligning MLM pre-training to ESPLADE fine-tuning. %
Also, the experimentation provides an empirical observation that, at most strict pruned settings, the retrieval effectiveness is better maintained by the higher-level retrieval cost, showing the trade-off relationship between the two in our setting. %

\section{Background}

\textbf{First-Stage Retrieval Methods in IR:}
In the field of Information Retrieval (IR), first-stage retrieval methods~(e.g., \citet{robertson2009probabilistic, luan2021sparse, formal2021splade}) are used directly by searchers~\citep{shah2022situating} as well as computer software like machine learning models~\citep{zamani2022retrieval, asai2024reliable}.
Recent research empirically shows that scaling the datastore of retrieval can be helpful for the performance of such machine learning models~\citep{shao2024scaling}, which is a similar trend to scaling the intrinsic factors of the model, such as parameter size or training data~\citep{fang2024scaling, kaplan2020scaling}.
For the past years, research on the first stage retrieval has developed to harness the representation powers of neural models, including dense vector retrieval~\citep{karpukhin2020dense, wang2022text}, or Learned Sparse Retrieval (LSR)~\citep{formal2021splade, formal2024towards, lassance2024two} with different emphases on how queries~(Q) and documents~(D) are represented.

\textbf{Learned Sparse Retrieval (LSR) and SPLADE Models:}
LSR has a characteristic that it can be served with inverted indexes~\citep{zobel2006inverted} instead of a similarity search on dense vector indexes~\citep{douze2024faiss}.
This characteristic promotes efficiency in serving and effectiveness in measuring relevance by concurrently integrating other features~\citep{liu2009learning} in inverted indexes at the level of first-stage matching.
SPLADE Models~\citep{formal2021splade, formal2024towards, lassance2024two} is an approach for end-to-end learning of sparse representations for ranking by utilizing the pooled result of MLM classification predictions of the BERT Model~\citep{devlin2019bert} with metric learning and FLOPS sparsity regularization loss~\citep{paria2020minimizing}.
The concept of SPLADE was further studied in terms of model architectures~\citep{zeng2025scaling, xu2025csplade}, output vocabularies~\citep{mackenzie2023exploring, yu2024improved, dudek2023learning, kim2025role}, static prunings~\citep{lassance2023static, lassance2024two, kim2025role}, regularization losses~\citep{dudek2023learning, porco2025alternative}, and pre-training~\citep{lassance2023experimental, yu2024improved, dudek2023learning, kim2025role}.

\textbf{Pre-training for SPLADE and Expanded-SPLADE:}
For the pre-training, \citet{lassance2023experimental, yu2024improved} conducted an analysis on pre-training of the SPLADE model from scratch and obtained a competitive performance. \citet{dudek2023learning} conducted a continual pre-training from existing pre-trained weights for a custom, expanded output vocabulary set, showing that after fine-tuning, retrieval effectiveness and efficiency are maintained.
Expanded-SPLADE model of \citep{dudek2023learning} replaces and uses a custom MLM layer (which is a fully connected layer) that leverages the weights and bias of the existing MLM layer.

For the additional backgrounds, a short survey of training neural language models for IR is presented in Appendix~\ref{a:short-ir-survey}.

\section{Pre-training, Fine-tuning, and Evaluation Methods}

Throughout this paper, we refer to the pre-trained model as EMLM (Expanded Masked Language Model) and the fine-tuned model as ESPLADE~(Expanded-SPLADE).
Both model was introduced by \citet{dudek2023learning}, and our procedure is based on that paper.

Here, we introduce their method at a high level.
The prediction label of the MLM task is a token of vocabulary.
The MLM outputs the probability of each token of the given vocabulary.
The specification of vocabulary is shared physically in the model by, the number of word embedding vectors in the input lookup table corresponding to each word in the vocabulary, and the weight vector of the output MLM FC layer for the prediction also corresponding to each word in the vocabulary.
Note that weight tying~\citep{press2017using} is often applied for sharing weight between them.

In the \citet{dudek2023learning}'s method, EMLM reconstructs the weight and bias of the output MLM FC layer, leveraging the original MLM FC layer. The new output vocabulary can have a different set of words, and a different total number of words~(or dimension).
Details of model construction, initialization,  pre-training masking strategy, and pre-training hyperparameters can be found in the Appendix \ref{emlm-model-details}.

\subsection{Pre-training Learning Rate and Dataset}
\label{sec313}

We pre-trained EMLM models with two learning rate configurations for each of three different datasets.

Pre-training learning rate configurations are:

\begin{itemize}
  \setlength\itemsep{0.5em}

\item \textit{*-lr-l.}~
This model is trained to 600K steps, and we used the last checkpoint for fine-tuning.
The learning rate decay method is linear.

\item \textit{*-lr-h.}~
This model is trained to 2.9M steps, and we use the 600K steps' checkpoint for fine-tuning.
The learning rate decay method is linear, hence models with this configuration were pre-trained with a higher "effective" learning rate compared to \textit{*-lr-l} models.

\end{itemize}

Pre-training dataset variations are:

\begin{itemize}
  \setlength\itemsep{0.5em}

\item \textit{emlm-ptd-overlap-repeat-lr-*.}~
This model has been trained with 614 million web titles, which are based on 77 million unique web titles from trainset-small with different masking patterns.
For the expanded vocabulary of this model, we have constructed the 100K most frequent unigrams from the trainset-small.
The pre-training corpus is overlapped with the fine-tuning corpus, since trainset-small is overlapped with trainset.
Note that the trainset-small and trainset are made for fine-tuning, however also used for pre-training. Please refer to the Appendix~\ref{a:finetuning-datasets} for the details.

\item \textit{emlm-ptd-indep-repeat-lr-*.}~
During training on the amount of 600K steps, the model sees a total of 614 million records, which are based on 79 million unique web titles from our in-house collection with different masking patterns. %
For the expanded vocabulary of this model, we have constructed the 100K most frequent unigrams from the trainset. This is to align the vocabulary's distribution to the trainset while pre-training on a different corpus.

\item \textit{emlm-ptd-indep-uniq-lr-*.}~
During training on the amount of 600K steps, the model sees a total of 614 million records, which are all unique web titles from our in-house collection. %
~
For the expanded vocabulary of this model, we used the same from \textit{emlm-ptd-indep-repeat-lr-*}.

\end{itemize}

Note that the two vocabulary sets are similar, with unigrams overlapping by 87.1\%
\footnote{
We expect the effect of vocabulary set difference to be non-significant for our study. Please refer to Appendix \ref{a:vocab-diff-details} for further details.
}
.

\subsection{Pre-training (EMLM) Validation Set}
\label{sec:emlm-validset}

We constructed an EMLM Validation Set (Validset) for each of the models' vocabulary.
We used 77,482 positive documents from the fine-tuning validset.
We choose the target unigram to be masked with the masking strategy described in the Appendix~\ref{a:pretraining-masking-strategy}.

Note that as the vocabulary of the two EMLM models is different, and the masking procedure involves a random selection process of the target token, the results validsets are not identical; hence, in this paper, we do not compare the results of different validset.

Each validset is comprised of two sets: "firstword-label-only" set, which only masks the first subword of WordPiece tokenized unigram, and "subword-label-only" set, which only masks all of the 2nd or later subwords of WordPiece tokenized unigram.

\subsection{Fine-tuning for ESPLADE Models}

We used part of our search log that contains users' query-document (Q-D) clicks. The detailed information for the dataset and training configuration of fine-tuning can be found in Appendix~\ref{a:ft-configs}.

\subsection{Q, D Pruning with MLM Logit Score}

We applied static term pruning \citep{lassance2023static} to documents as well as queries based on the MLM logit score to max size. This decision was made to examine learned sparse representations that balance retrieval effectiveness and efficiency.
More background and details can be found in Appendix \ref{a:q-d-pruning}.

\subsection{Evaluation Set}

\begin{table}
    \centering

\small
  \begin{tabular}{c|r}
    \toprule
    Document Count & 20,372,952 \\
    \midrule
    Test Query Count & 8,936 \\
    \midrule
    Avg. Doc. Title Length & 13.4 words \\
    \midrule
    Avg. Query Length & 2.6 words \\
  \bottomrule
\end{tabular}
\caption{\textbf{Stats for Evaluation Set.}}
  \label{tab:eval-set}

\end{table}

The evaluation set is constructed to measure retrieval effectiveness and efficiency.
Table \ref{tab:eval-set} shows the stats of the evaluation set.

\subsubsection{Sampling Queries}
We sampled triplets of query, positive document list, and negative document list from the log.
To select the evaluation query set $Q$, we cluster queries using an embedding model.
We use this clustering result to pick candidate queries that are more semantically balanced.
We sample these candidate queries to construct the $Q$.
We picking up queries based on the matched term ratios to their corresponding positive documents (titles). Note that among matched term ratios of multiple positive documents, we use the highest.
Based on that value, we classify queries into three quantiles - low, middle, and high matched term ratios.
We pick the same number of queries in each quantile.
This was performed to obtain queries that have a diverse level of lexical matching difficulty.

\subsubsection{Sampling Positive and Negative Documents for Query}

For positives, we select the top-k positive documents based on label score. For each query, we pick 3 documents at most.
For negatives, we select negative documents for each target query from positive documents of queries in other clusters that above a fixed value of high term matching ratio to the target query.
(i.e., naive lexical matching hurts the performance).
For each target query, we obtain negative documents by selecting the top 10 documents at most that have the highest term matching ratio to the target query.
We only use triplets where all query, positive documents, and negative documents exist.

\subsection{Retrieval Efficiency Measure}
To evaluate retrieval efficiency, we define the FLOPS metric that approximates the volume of 1st stage matched documents in the search engine by combining of postings list lengths for each query in the evaluation query set.

This FLOPS metric measures the computational cost of retrieval on a search engine, as it approximates the traversal cost of the postings list, which reflects the traversal cost itself, and the cost of more expensive later-stage ranking proportional to the volume of first matched documents. The FLOPS metric is defined as:

\begin{equation}
\text{FLOPS} = \frac{\sum_{q \in Q} \sum_{t \in q} \left| P_t \right|}{|Q| \cdot |D|}
\end{equation}

Where $Q$ is the set of evaluation queries, $t$ is a query term in the individual evaluation query.
$|P_t|$ is the postings list length, which is the number of matching documents belonging to term $t$.
$|Q|$ and $|D|$ are the total count of queries and documents. %

\section{Experimentation Results}

\subsection{Model Performance}

\subsubsection{Metrics for EMLM Logit Vectors}
\label{sec:emlm-logit-vector}

To characterize distributions of scores in the EMLM logit vectors, we define a statistic, logit-score-std (an abbreviation of Logit-wise Score Std), that is the average of the standard deviation of individual logit indices’ scores appearing in logit vectors from documents.
We used logit vectors of 164,740 WordPiece tokens from 8,000 positive documents from validset, where each logit vector has 100K dimensions.
This statistic can characterize part of the model's fitness to the training objective of MLM classification loss.
Appendix~\ref{a:statistics-emlm-word-embeddings} shows more statistics of EMLM logit vectors.

\subsubsection{Pre-training Configurations and Retrieval Effectiveness}

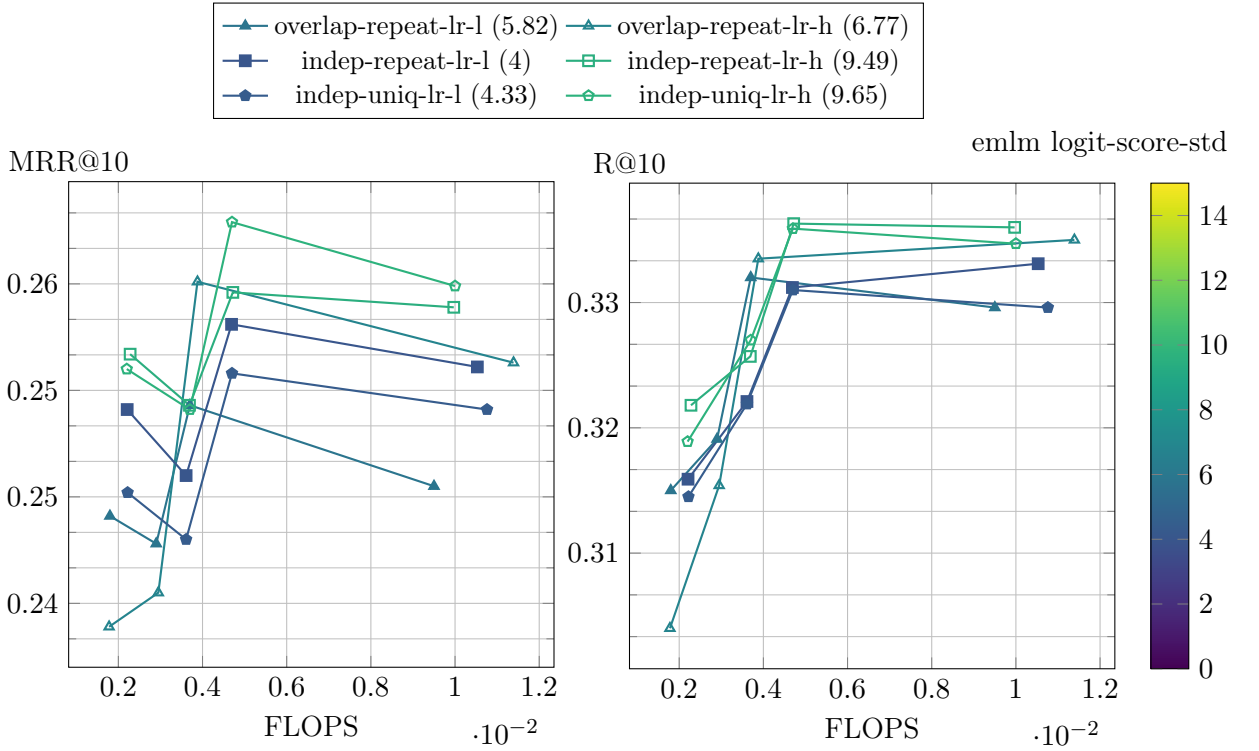
\begin{figure*}

\centering{
\pgfplotslegendfromname{sharedlegend3}
}

\begin{minipage}{0.46\textwidth} %
\vspace{7pt}

                \begin{tikzpicture}
            \begin{axis}[
                xlabel={FLOPS},
                ylabel={MRR@10}, %
                xlabel near ticks,
                ylabel style={at={(0,1)}, anchor=south,rotate=-90},
                grid=major,
                width=8cm, %
                height=8cm,%
                legend style={font=\small},
                legend to name={sharedlegend3}, %
                legend columns=2, %
                minor y tick num=2,  %
                minor grid style={lightgray,very thin},
                grid=both,
            ]

                \addplot[color={rgb,255: red,43; green,117; blue,142}, mark=triangle*, thick] coordinates {

(0.0095,  0.2455)
(0.0037,  0.2493)
(0.0029,  0.2428)
(0.0018,  0.2441)

                };
                \addlegendentry{overlap-repeat-lr-l (5.82)}

                \addplot[color={rgb,255: red,37; green,132; blue,142}, mark=triangle, thick] coordinates {

(0.01139, 0.2513)
(0.00388, 0.2551)
(0.00295, 0.2405)
(0.00178, 0.2389)

                };
                \addlegendentry{overlap-repeat-lr-h (6.77)}

                \addplot[color={rgb,255: red,57; green,86; blue,140}, mark=square*, thick] coordinates {

(0.01053, 0.2511)
(0.00469, 0.2531)
(0.00361, 0.246)
(0.00221, 0.2491)

                };
                \addlegendentry{indep-repeat-lr-l (4)}

                \addplot[color={rgb,255: red,43; green,174; blue,127}, mark=square, thick] coordinates {

(0.00997, 0.2539)
(0.00472, 0.2546)
(0.00369, 0.2493)
(0.00228, 0.2517)

                };
                \addlegendentry{indep-repeat-lr-h (9.49)}

                \addplot[color={rgb,255: red,54; green,92; blue,141}, mark=pentagon*, thick] coordinates {

(0.01076, 0.2491)
(0.0047,  0.2508)
(0.00361, 0.243)
(0.00222, 0.2452)

                };
                \addlegendentry{indep-uniq-lr-l (4.33)}

                \addplot[color={rgb,255: red,45; green,178; blue,125}, mark=pentagon, thick] coordinates {

(0.01,    0.2549)
(0.0047,  0.2579)
(0.0037,  0.2491)
(0.0022,  0.251)

                };
                \addlegendentry{indep-uniq-lr-h (9.65)}

            \end{axis}
        \end{tikzpicture}
\end{minipage}
\begin{minipage}{0.01\textwidth}
~
\end{minipage}
\begin{minipage}{0.46\textwidth}
                       \begin{tikzpicture}
            \begin{axis}[
                xlabel={FLOPS},
                ylabel={R@10}, %
                xlabel near ticks,
                ylabel style={at={(0,1)}, anchor=south,rotate=-90},
                grid=major,
                width=8cm, %
                height=8cm,%
                legend to name={sharedlegenddummy}, %
                minor y tick num=2,  %
                minor grid style={lightgray,very thin},
                grid=both,
                colorbar,
                colorbar style={
                    title={emlm logit-score-std},
                    title style={at={(-1.35,1.0)}},
                    },
                colormap name=viridis,
                point meta min=0,
                point meta max=15,
            ]

                \addplot[color={rgb,255: red,43; green,117; blue,142}, mark=triangle*, thick] coordinates {

(0.0095,  0.3296)
(0.0037,  0.332)
(0.0029,  0.3191)
(0.0018,  0.315)

                };
                \addlegendentry{overlap-repeat-lr-l (5.82)}

                \addplot[color={rgb,255: red,37; green,132; blue,142}, mark=triangle, thick] coordinates {

(0.01139, 0.335)
(0.00388, 0.3335)
(0.00295, 0.3154)
(0.00178, 0.304)

                };
                \addlegendentry{overlap-repeat-lr-h (6.77)}

                \addplot[color={rgb,255: red,57; green,86; blue,140}, mark=square*, thick] coordinates {

(0.01053, 0.3331)
(0.00469, 0.3312)
(0.00361, 0.3221)
(0.00221, 0.3159)

                };
                \addlegendentry{indep-repeat-lr-l (4)}

                \addplot[color={rgb,255: red,43; green,174; blue,127}, mark=square, thick] coordinates {

(0.00997, 0.336)
(0.00472, 0.3363)
(0.00369, 0.3257)
(0.00228, 0.3218)

                };
                \addlegendentry{indep-repeat-lr-h (9.49)}

                \addplot[color={rgb,255: red,54; green,92; blue,141}, mark=pentagon*, thick] coordinates {

(0.01076, 0.3296)
(0.0047,  0.331)
(0.00361, 0.3219)
(0.00222, 0.3145)

                };
                \addlegendentry{indep-uniq-lr-l (4.33)}

                \addplot[color={rgb,255: red,45; green,178; blue,125}, mark=pentagon, thick] coordinates {

(0.01,    0.3347)
(0.0047,  0.3359)
(0.0037,  0.327)
(0.0022,  0.3189)

                };
                \addlegendentry{indep-uniq-lr-h (9.65)}

            \end{axis}
        \end{tikzpicture}
\end{minipage}
\begin{minipage}{0.045\textwidth}
~
\end{minipage}

\caption{\textbf{Result on Evaluation Set of Fine-tuned Models with Different Pre-trained EMLM Models}
\small{
Models are trained on trainset (see Appendix \ref{a:finetuning-datasets} for the details), with top-k masking~\citep{yang2021sparsifying} of q\_K=1000, d\_K=2000.
Q, D pruning (see Appendix \ref{a:q-d-pruning} for the detailed explanation) is applied for the rightmost node to the leftmost node (qk=0, dk=0; not pruned), (qk=7, dk=20), (qk=5, dk=20), (qk=5, dk=10), respectively.
The logit-score-std value of each pre-trained EMLM model is associated with the name of the corresponding fine-tuned model in the legend. The value is also used to color each line of the graph.
The validset result is presented in the Table \ref{tab:validset-pref} of the Appendix \ref{a:hps}.
Table~\ref{finetuning-pref-table} is the table of this figure.
}}
  \label{finetuning-pref-figure}
\end{figure*}

The Figure~\ref{finetuning-pref-figure} shows fine-tuning performance.
Models with a higher learning rate (\textit{*-lr-h}) show a higher logit-score-std score in their pre-trained model.
Among them, models pre-trained on general datasets (\textit{indep-*}) show mostly higher retrieval effectiveness on their fine-tuned models in both unpruned and most strict pruning settings.
Where another one, pre-trained on an in-fine-tuning dataset (\textit{overlap-*}), shows mostly lower retrieval effectiveness on their fine-tuned models in both unpruned and most strict pruning settings compared to other \textit{*-lr-h} models.

\subsubsection{Retrieval Effectiveness and Efficiency}

\begin{table*}
\small
    \centering

  \setlength\tabcolsep{2pt}
  \begin{tabular}{lS[table-format=1.2]ccS[table-format=2.2]S[table-format=2.2]S[table-format=1.5]S[table-format=1.4]S[table-format=1.4]}
    \toprule

model & {emlm logit-score-std} & qk & dk & {L0\_q} & {L0\_d} & {FLOPS} & {MRR@10} & {R@10}  \\
\midrule  
bm25 & . & . & . & . & . & 0.00277 & 0.203 & 0.2527  \\
\midrule  
overlap-repeat-lr-l & 5.82 & 0 & 0 & 11.29 & 55.36 & {\sibf{0.0095}} & 0.2455 & 0.3296  \\
overlap-repeat-lr-h & 6.77 & 0 & 0 & 13.05 & 58.42 & 0.01139 & 0.2513 & {\siuline{0.335}} \\
indep-repeat-lr-l & 4 & 0 & 0 & 10.39 & 61.14 & 0.01053 & 0.2511 & 0.3331  \\
indep-repeat-lr-h & 9.49 & 0 & 0 & {\sibf{9.58}} & {\siuline{45.73}} & {\siuline{0.009 97}} & {\siuline{0.2539}} & {\sibf{0.336}}  \\
indep-uniq-lr-l & 4.33 & 0 & 0 & 10.48 & 63.51 & 0.01076 & 0.2491 & 0.3296  \\
indep-uniq-lr-h & 9.65 & 0 & 0 & {\siuline{9.77}} & {\sibf{39.9}} & 0.01 & {\sibf{0.2549}} & 0.3347  \\
\midrule  
overlap-repeat-lr-l & 5.82 & 7 & 20 & 6.55 & 19.23 & {\sibf{0.0037}} & 0.2493 & 0.332  \\
overlap-repeat-lr-h & 6.77 & 7 & 20 & 6.76 & 19.48 & {\siuline{0.003 88}} & {\siuline{0.2551}} & 0.3335  \\
indep-repeat-lr-l & 4 & 7 & 20 & 6.45 & {\siuline{18.59}} & 0.00469 & 0.2531 & 0.3312  \\
indep-repeat-lr-h & 9.49 & 7 & 20 & {\sibf{6.29}} & 18.91 & 0.00472 & 0.2546 & {\sibf{0.3363}}  \\
indep-uniq-lr-l & 4.33 & 7 & 20 & 6.46 & 18.73 & 0.0047 & 0.2508 & 0.331  \\
indep-uniq-lr-h & 9.65 & 7 & 20 & {\siuline{6.31}} & {\sibf{18.29}} & 0.0047 & {\sibf{0.2579}} & {\siuline{0.3359}}  \\
\midrule  
overlap-repeat-lr-l & 5.82 & 5 & 20 & 4.92 & 19.23 & {\sibf{0.0029}} & 0.2428 & 0.3191  \\
overlap-repeat-lr-h & 6.77 & 5 & 20 & 4.96 & 19.48 & {\siuline{0.002 95}} & 0.2405 & 0.3154  \\
indep-repeat-lr-l & 4 & 5 & 20 & 4.89 & {\siuline{18.59}} & 0.00361 & 0.246 & 0.3221  \\
indep-repeat-lr-h & 9.49 & 5 & 20 & {\sibf{4.84}} & 18.91 & 0.00369 & {\sibf{0.2493}} & {\siuline{0.3257}}  \\
indep-uniq-lr-l & 4.33 & 5 & 20 & 4.89 & 18.73 & 0.00361 & 0.243 & 0.3219  \\
indep-uniq-lr-h & 9.65 & 5 & 20 & {\sibf{4.84}} & {\sibf{18.29}} & 0.0037 & {\siuline{0.2491}} & {\sibf{0.327}}  \\
\midrule  
overlap-repeat-lr-l & 5.82 & 5 & 10 & 4.92 & 9.96 & {\siuline{0.0018}} & 0.2441 & 0.315  \\
overlap-repeat-lr-h & 6.77 & 5 & 10 & 4.96 & 9.98 & {\sibf{0.001 78}} & 0.2389 & 0.304  \\
indep-repeat-lr-l & 4 & 5 & 10 & 4.89 & {\siuline{9.92}} & 0.00221 & 0.2491 & 0.3159  \\
indep-repeat-lr-h & 9.49 & 5 & 10 & {\sibf{4.84}} & 9.94 & 0.00228 & {\sibf{0.2517}} & {\sibf{0.3218}}  \\
indep-uniq-lr-l & 4.33 & 5 & 10 & 4.89 & 9.93 & 0.00222 & 0.2452 & 0.3145  \\
indep-uniq-lr-h & 9.65 & 5 & 10 & {\sibf{4.84}} & {\sibf{9.91}} & 0.0022 & {\siuline{0.251}} & {\siuline{0.3189}}  \\

  \bottomrule
\end{tabular}

\caption{\textbf{Result on Evaluation Set of Fine-tuned Models with Different Pre-trained EMLM Models}
\small{
For the qk and dk, the value of 0 means unpruned.
L0\_q, L0\_d is the average number of Q, D terms used for the evaluation after the pruning is applied.
In each pruning setting, for each metric, the best metric value and the 2nd best metric value are highlighted in bold and underlined, respectively.
More description is included in the caption of Figure~\ref{finetuning-pref-figure}.
}}
  \label{finetuning-pref-table}
\end{table*}

The Table~\ref{finetuning-pref-table} shows that in strict pruned settings of (qk=5, dk=20) and (qk=5, dk=10), \textit{indep-*} models show generally higher MRR and R performance as well as higher FLOPS.
A trade-off relationship between retrieval effectiveness and efficiency can be observed.

\subsubsection{Postings List Length Distributions and Retrieval Effectiveness, Efficiency}
\label{pos-ls-len-dist-and-retrieval-ee}

\begin{table*}
\centering
\footnotesize
  \begin{tabular}{l|ccc}
    \toprule
esplade model and pruning strategies & mean & var & std\\
\midrule
overlap-repeat-lr-l-d\_k0 & 24.54 & 1522 & 39.01\\
indep-repeat-lr-l-d\_k0 & 22.08 & 1438 & 37.92\\
indep-uniq-lr-l-d\_k0 & 23.07 & 1441 & 37.96\\
\midrule
overlap-repeat-lr-h-d\_k0 & 28.68 & 1607 & 40.09\\
indep-repeat-lr-h-d\_k0 & 21.14 & 1482 & 38.49\\
indep-uniq-lr-h-d\_k0 & 19.5 & 1484 & 38.53\\
\midrule
overlap-repeat-lr-l-d\_k20 & 13.93 & 908 & 30.14\\
indep-repeat-lr-l-d\_k20 & 13.21 & 958 & 30.96\\
indep-uniq-lr-l-d\_k20 & 13.35 & 958 & 30.95\\
\midrule
overlap-repeat-lr-h-d\_k20 & 14.48 & 808 & 28.42\\
indep-repeat-lr-h-d\_k20 & 13.44 & 947 & 30.77\\
indep-uniq-lr-h-d\_k20 & 13.03 & 959 & 30.97\\
\midrule
overlap-repeat-lr-l-d\_k10 & 7.62 & 369 & 19.22\\
indep-repeat-lr-l-d\_k10 & 7.56 & 396 & 19.89\\
indep-uniq-lr-l-d\_k10 & 7.57 & 393 & 19.83\\
\midrule
overlap-repeat-lr-h-d\_k10 & 7.68 & 305 & 17.46\\
indep-repeat-lr-h-d\_k10 & 7.56 & 402 & 20.06\\
indep-uniq-lr-h-d\_k10 & 7.54 & 397 & 19.92\\

  \bottomrule
\end{tabular}
  \caption{\textbf{Stats of the individual postings list length for each model.} Target data is 77,482 positive document titles of the validset. d\_k is the remaining top-k terms after pruning, where d\_k0 means unpruned.}
  \label{tab:402}
\end{table*}

The Table \ref{tab:402} displays the mean, variance, and standard deviation~(std) of the length of the individual postings list with different pruning strategies.
In this table, a higher std value means the length of the postings list for individual terms is distributed in a high-variance (less uniform) manner.

In an unpruned setting, the model of comparably moderate retrieval effectiveness (\textit{overlap-repeat-*}) shows a higher std value in their postings list length compared to its counterparts.
In strict pruned settings of (qk=5, dk=20) and (qk=5, dk=10), models of comparably higher retrieval effectiveness (\textit{indep-*}) show a higher std value in their postings list length compared to each of their counterpart models.

The \textit{indep-repeat-lr-h-d\_k10} and \textit{indep-uniq-lr-h-d\_k10} models are the two models that show higher retrieval effectiveness with higher-level FLOPS in the most strict pruned settings (shown in Table \ref{finetuning-pref-table}). These models show a higher std value in the length of the individual postings list (i.e., produce more common terms from different text records) compared to the \textit{overlap-repeat-lr-h-d\_k10} model, as shown in Table \ref{tab:402}\footnote{Please check Appendix~\ref{a:pos-ls-len-var-and-retrieval-ee} for the additional discussion.}.

Overall, this result reflects a case of trend between retrieval effectiveness, efficiency, and the degree of uniformity of postings list length in a strict pruning setting, on the basis of differences in pre-training.
One trend we observed is a trade-off relationship between effectiveness and efficiency in a strict pruning setting.
The variance in the length of the individual postings list originated from the differences in pre-training can be related.

\subsection{Retrieval Effectiveness and Pre-training Accuracies, Losses}

\input{asset-8-emlm-pretraining-metrics}

The Figure~\ref{fig:emlm-pretraining-accuracies} shows accuracies of the EMLM task during pre-training using validset of Section~\ref{sec:emlm-validset}.
The two comparably well-performing EMLM models for retrieval effectiveness when fine-tuned are emlm-ptd-indep-repeat-lr-h and emlm-ptd-indep-uniq-lr-h, which correspond to the indep-repeat-lr-h and indep-uniq-lr-h, respectively.
These models have a higher EMLM logit-score-std score and show lower accuracies.
The Figure~\ref{fig:emlm-pretraining-losses} shows pre-training losses, which follows the inverse trend of accuracy.
These provide a case that improving EMLM accuracies are not always aligned with the retrieval effectiveness of SPLADE fine-tuning~
\footnote{In addition, some longer steps pre-trained checkpoints that have higher accuracies lead to the failure of fine-tuning. See Appendix~\ref{apdx:pretraining-steps-effect} for the details.}.

\section{Discussion}

\subsection{Adaptability of EMLM Pre-training for Expanded-SPLADE Fine-tuning}

In this paper, we studied the pre-training of Expanded-SPLADE models on web document titles.
In the main experimentation result of Figure~\ref{finetuning-pref-figure}, we found cases where higher prediction accuracies of the EMLM task or pre-training with an in-fine-tuning dataset can not be an effective factor for fine-tuning Expanded-SPLADE models.
The experimental result shares partly similar directions to previous work that suggest the MLM pre-training and the task of IR are not always aligned~\citep{gao2021condenser, lassance2023experimental}.
This identifies the needs of new pre-training methods for LSR consulted for such alignment.
We expect pre-training task of inter-sentence discrimination, instead of intra-sentence predication of MLM, may be needed.
Pre-training methods studied specifically for the retrieval task are developed, as shown in the Appendix~\ref{a:pt-and-ir}.
However, approaches for pre-training high-dimensional output, which can be used for fine-tuning LSR, have been studied limitedly.

Models pre-trained with the general corpus can have higher logit-score-std scores and show high performance when fine-tuned.
With this, we can speculate that in the pre-training with general corpus, the adaptability of pre-trained models for Expanded-SPALDE fine-tuning can be related to the logit-score-std score of pre-trained models.

We assume that fine-tuning for ranking is a different objective from fine-tuning for classification.
As ranking involves similarity between queries and documents, different from MLM, which maximizes the predictive powers of a single instance by self-supervision \citep{devlin2019bert}
\footnote{
The pre-training loss is MLM loss, an intra-instance(a sentence in our case) prediction task,
and the fine-tuning loss of SPLADE is typically a composite of two losses, a contrastive loss and FLOPS regularization loss, which corresponds to the effectiveness and efficiency metric or criterion of retrieval. These fine-tuning losses are related to metric learning, which is an inherently task of inter-instance discrimination.
}
\footnote{Naive alternative approach can be a pre-training task that uses the same classification loss for MLM, but involves the whole context of query and document, such as concatenated Q and D as an MLM input.
This may improve the neutralness of representations in terms of input data; however, discrepancies can still exist in terms of performing fine-tuning metric learning, starting from the same representations derived from the 'concatenated' intra-sentence prediction task, to learning latent representations for Q, D, which have asymmetric lexical and semantic differences.
 }.
Overall, this provides evidence of the discrepancy between EMLM pre-training and Expanded-SPLADE fine-tuning.
In this context, the main experimentation result of Figure~\ref{finetuning-pref-figure} can be interpretable as,
the better adaptable pre-trained models for ESPLADE fine-tuning (i.e., having smaller discrepancy between EMLM pre-training and Expanded-SPLADE fine-tuning) demonstrate higher retrieval effectiveness when both training-time regularization (FLOPS regularization) and test-time efficiency technique (static pruning) are applied.

\subsection{The Repetition of Pre-training Dataset}

From the main experimentation result of Figure~\ref{finetuning-pref-figure}, we observed that the repetition of the general pre-training dataset does not have much effect on retrieval effectiveness.
Among the models of indep-repeat-lr-h and indep-uniq-lr-h, the indep-repeat-lr-h shows slightly higher retrieval effectiveness when the most strict pruning is applied.
This may be another evidence that the diversity of the corpus for the MLM classification task is not crucial for ranking fine-tuning with FLOPS regularization, which implies the limitation of transferring (or perhaps necessity of transferring) contextual knowledge from the MLM pre-training corpus to ranking fine-tuning, which is aligned with the experimentation from \citet{lassance2023experimental}.
However, to the best of our knowledge, in BERT models for ESPLADE, we currently have limited knowledge that pre-training methods other than MLM can harness a larger corpus, and make differences in the retrieval effectiveness of fine-tuned models. Finding such pre-training methods can explore pre-training scaling laws for retrieval on ESPLADE models. Appendix~\ref{scaling-laws} shares some related works.

\section{Conclusion}

In this paper, we studied pre-training for ESPLADE models on web document titles.
The three observations and findings of our study are as follows.
First, we have observed that fine-tuned models of higher retrieval effectiveness at both unpruned and most strict pruned settings are mostly pre-trained on a general corpus, and pre-trained with a higher learning rate, showing lower MLM accuracies with higher score variance in each logit index.
The first finding provides a case where ESPLADE fine-tuning does not always align with MLM pre-training, as well as pre-training with the in-fine-tuning corpus.
Second, we observed that in the most strict pruned setting, those models show higher-level retrieval cost and a higher variance in the length of the individual postings list.
The second finding provides a case that shows how the effectiveness-efficiency trade-off is implemented in the level of the postings list when strict pruning is applied.
Third, we observed that the repetition of the general pre-training dataset does not have much effect on retrieval effectiveness. %
The third finding provides a case of the limitation (or perhaps necessity) of transferring contextual knowledge from the diverse MLM pre-training corpus to ranking fine-tuning of ESPLADE models.

\bibliography{custom}

\acks{
We thank members of the NAVER Search team for their support.
We appreciate anonymous reviewers of IRRJ for providing comments to our SPLADE related research.
}

\clearpage

\appendix

\section{A Short Survey of Training Neural Language Models for Information Retrieval}
\label{a:short-ir-survey}

\subsection{Pre-Training and Information Retrieval}
\label{a:pt-and-ir}

\citet{erhan2010does} suggests pre-training can provide regularization, leading the training of the network towards better generalization of the downstream training dataset.

Pre-training approach for training representation learning models has been widely adopted for the sake of academic and research environment of larger-scale datasets, effective off-the-shelf model architectures, and performant accelerators, with various modalities and tasks~\citep{han2021pre}.
Natural language processing~(NLP) is one of the areas affected by pre-training.

BERT~\citep{devlin2019bert} is a pre-trained transformer~\citep{vaswani2017attention} encoder, has widely used for IR~\citep{lin2022pretrained}, as a form of bi-encoder~\citep{karpukhin2020dense, khattab2020colbert, formal2021splade} or cross-encoder~\citep{nogueira2019passage} models.

GPT~(e.g., \citet{radford2018improving, radford2019language}) is a pre-trained transformer decoder, whose architectural concepts are also employed for bi-encoder-based IR tasks~\citep{springer2024repetition, suganthan2025adapting, lee2025gemini, nie2024text}.

Aside from this, pre-training methods were studied specifically for the retrieval task, which involves the concept of similarity.
\citet{lee2019latent} presents the Inverse Cloze Task (ICT), given the span of a sentence, learning to increase the similarity of neighboring sentences, teaching the model to capture semantic relationships that can not be revealed in a lexically similar relationship, shown to be effective in the retrieval setting~\citep{chang2020pre}.

Similar to ICT, \citet{izacard2021unsupervised} presents Contriever, which independently crops two spans of a document to obtain positive pairs and conduct contrastive learning.
This is similar to self-supervised pre-training from computer vision models~\citep{caron2021emerging} that uses croppings of the same image~\citep{caron2020unsupervised} as a positive pair, instead of conducting pre-training with masked patch prediction on images~\citep{dosovitskiy2020image}.
In addition, Contriever employs MoCo~\citep{he2020momentum} technique, which uses a queue to store precomputed negatives from the slowly updated encoder regulated by momentum update, allowing the model to learning from a large number of negative examples.
Similar contrastive pre-training methods have been extensively studied~\citep{neelakantan2022text, wang2022text}.

\citet{gao2021condenser} presents Condenser, a modified BERT architecture for pre-training to conditioning the BERT model fits to dense retrieval. Specifically, they remove connections to the head and backbone layer other than the CLS token and some short circuit path from the early layer, enforce the model to compress information in the CLS token, while alleviating the need to encode all local and lexical information to that token by employing a short circuit path from the early layer.
Note that they initialize their backbone layers from the pre-trained BERT model, and head layers are initialized randomly.
On top of this architecture, they use MLM loss for the pre-training.
They also empirically suggest that the structural readiness of typical transformer encoders is limited to be fine-tuned into bi-encoders, need a process of large internal structural changes, while Condenser and ICT can have less such discrepancy.

Note that many of these models are not completely pre-training from scratch~(like~\citet{lassance2023experimental}), Contriever models~\citep{izacard2021unsupervised} are obtained after continual pre-training based on a pre-trained BERT backbone. \citet{lee2019latent, wang2022text, gao2021condenser} shares a similar setup in their paper. \citet{neelakantan2022text} uses a pre-trained GPT backbone.
In contrast to this, the experimentation of \citet{chang2020pre} appears to involve pre-training from scratch with their designated pre-training tasks. In addition, their experimentation includes no pre-training setup, which represents random initializing the model for fine-tuning.

In this paper, we focus on the continual pre-training approach.
Instead of studying the pre-training methods for dense retrieval, we are focusing on pre-training and predictions of MLM for LSR.

\subsection{Study on the Effect of Pre-training for Learned Sparse Retrieval}
\citet{lassance2023experimental} shows that pre-training with the IR fine-tuning dataset, which is smaller than the general-purpose pre-training dataset, does not negatively affect downstream ranking performance.
Following the empirical evidence of this work, \citet{yu2024improved} presents an approach to improve the effectiveness and efficiency of LSR models by constructing WordPiece vocabulary from the target corpus and pre-training the BERT from scratch on the target corpus.

ESPLADE \cite{dudek2023learning} shows an approach to expand vocabulary for an arbitrary set of vocabulary using a pre-training task called expanded masked language modeling.
This approach leverages the pre-trained weights of WordPiece-based plain BERT backbones to construct a new vocabulary with different semantics and thus can have a different size (dimension).
The study shows the ESPLADE models maintain effectiveness and efficiency compared to the original SPLADE models.

\citet{mackenzie2023exploring} assesses the representation powers of SPLADE models with different controlled vocabularies.
Specifically, they discuss that the size of vocabulary and the pre-trained weight of the MLM head are important factors in the fine-tuning performance of SPLADE models.

The aforementioned three approaches conduct 1) pre-training from scratch, 2) leverage existing pre-trained weights for continual pre-training, and 3) comparison of pre-trained MLM head weights versus randomly initialized MLM head weights and experimentation regarding vocabulary size, respectively.
While these studies involve models with MLM pre-training, observational examinations of the role and limits of MLM are not much discussed, which is one of the goals of this paper.

\subsection{Sparsification and Retrieval Efficiency of Learned Sparse Representations}

SPLADE \citep{formal2021splade} models employ the FLOPS regularization loss \citep{paria2020minimizing} to sparsify query and document vectors while minimizing the ranking loss. The ESPLADE models \citep{dudek2023learning} propose a joint FLOPS regularization loss, which sparsifies the intersection of query and document, which are more directly related to the retrieval efficiency.
\citep{yang2021sparsifying} presents ways to learning more sparsified representations by fine-tuning with modified pooled BERT logits that only use the top-k highest scores along with original pooled BERT logits.

In the test time, \citet{lassance2023static} presents that applying term pruning techniques for SPLADE-based inverted indexes is helpful for efficient retrieval with marginal sacrifice on the effectiveness.

\subsection{Scaling Laws in Neural Language Models}
\label{scaling-laws}
\citet{kaplan2020scaling} discusses scaling laws of neural language models, which are predictable patterns of model performance to training configurations such as model size, dataset size, and amount of compute.
Notably, they show that, in the fixed model size, larger dataset size decreases test loss, along with the observation of model size needs to be increased to fully utilize the larger dataset size.
However, this study focuses on decoder-only transformer architecture, also known as causal language models, which is different from masked language modeling that is currently commonly used for LSR.

\citet{fang2024scaling} studied scaling laws for dense retrieval, attempting to examine predictable patterns of model size and fine-tuning dataset size to ranking performance.
Specifically, they show that, in the fixed model size, increasing the fine-tuning dataset size is helpful for ranking performance.
Our study involves the effect of the MLM pre-training and its dataset on LSR.

\clearpage

\section{EMLM Model Details}
\label{emlm-model-details}

\subsection{Model Construction and Initialization}

We used an in-house BERT model that has 6 layers, 768 hidden dimensions, 12 heads, 55M parameters, and a vocabulary size of 32001. The model is pre-trained from corpora that include news, books, wikis, and encyclopedias centric primarily to Korean. The BERT model uses a subword tokenizer. We set the maximum length accepted by the model to 64.

We refer to the procedure on \cite{dudek2023learning} to construct and initialize EMLM models.
Here we describe the procedure for illustrative purposes.
First, given the corpus, we split the corpus into a train and a validation set that are exclusive.
Then get the top 100K frequentist vocabulary after extracting all of the unigrams in the corpus and sorting by their frequency.
Note that each vocabulary must be tokenized into one or more WordPiece subwords.
This vocabulary is the expanded vocabulary set $U$.
We construct the weight and bias of the EMLM head by performing subword tokenization on each vocabulary word $U$.
For instance, the term "love" can be compounded by two WordPiece subword vocabulary "lo" and "\#\#ve", then we mean pooling the two words' weight and bias from the MLM FC head to construct the weight and bias of expanded vocabulary "love" in the $U$.
We replace the MLM head layer with a newly constructed weight and bias of 100K $U$ terms.

\subsection{Pre-training Masking Strategy}
\label{a:pretraining-masking-strategy}

We mask $U$ vocabulary terms by the ratio of 15\% in each title.
For instance, a sentence that has 20 $U$ terms, we mask and only mask WordPiece terms corresponding to the selected 3 $U$ terms.
Note that this masking strategy is different from conducting the Bernoulli trial for each $U$ term by 15\%.
We guarantee at least 1 $U$ term to be masked if the count of 15\% $U$ terms is below 1 (e.g., short title).
We follow the masking strategies details of the original BERT paper \citep{devlin2019bert}, 80\% of target mask tokens will be transformed into \texttt{[MASK]} token, 10\% random token, 10\% unchanged.
If the target mask $U$ token consists of multiple WordPiece tokens, we mask all subwords (i.e., mask the first and all later-positioned subwords).
The target mask token will have the label of the corresponding $U$ token.

\subsection{Pre-training Hyperparameters}

We used the learning rate of 1.6e-4 with linear learning rate scheduling.
Please refer to Section~\ref{sec313} of the main text for the training steps and their impact on the learning rate.
We used the warmup of 10,000 steps.
The batch size is 1024 with 8 A100 GPUs.
We used AdamW optimizer with weight decay of 0.01.
We used 16-bit (mixed) precision training.

\clearpage

\section{Fine-tuning Details}
\label{a:finetuning-details}

\begin{table*}[ht]
    \centering

\footnotesize
\begin{tabular}{lrrrrr}
\toprule
dataset & \# of q-title pairs & \# of unique titles & \# of queries & docs per query & queries per doc\\
\midrule
trainset & 262,959,262 & 114,139,846 & 34,145,679 & 7.701 & 2.304\\
trainset-small & 182,784,463 & 79,154,465 & 21,394,466 & 8.543 & 2.104\\
testset & 9,629,311 & 8,210,573 & 1,198,184 & 8.036 & 1.070\\
validset & 339,851 & 307,901 & 39,998 & 8.496 & 1.004\\
\bottomrule
\end{tabular}
\caption{\textbf{Stats for Finetuning Datasets.} \small{validset and testset include distractor queries and documents.}}
\label{tab:stats-set}

\end{table*}

\subsection{Fine-tuning Datasets}
\label{a:finetuning-datasets}

The model was trained on a dataset of queries and web document titles.
We used part of our search log that contains users' query-document (Q-D) clicks\footnote{Due to the diverse nature of queries and documents that we possess, the dataset is multilingual, but centric primarily to Korean. There have been concepts in quantitative linguistics, such as Zipf's law, and recent work (e.g., \citet{zoph2016transfer, johnson2017google, feng2020language}) in the NLP community identifies the similarity in the latent structure of different languages, by emphasizing the transferability between different languages in NLP tasks. Although further experimentation is fruitful, we foresee the findings of this paper can be applicable to other languages as well.}.

Table \ref{tab:stats-set} shows stats for the dataset.
To obtain a clear supervision signal, we use only when there are multiple documents per query that appear within the upper part of the search results and have been clicked on multiple times. 
We allow the same document title to appear multiple times in the dataset.
For constructing the testset and validset, we use our in-house SBERT-based sentence embedding model to cluster these datasets into 10K clusters based on the semantics of queries.

\subsubsection{Testset}
Given the clustered result, we randomly pick 30 queries from each cluster (a total of 300K queries) and use these queries and associated documents as labels.
For the distractor, we randomly pick  90 queries from each cluster and get their documents. (documents from a total of 900K queries).
We exclude documents that have already been picked as positive for a distractor.

\subsubsection{Validset}
Given the clustered result, we randomly pick 1 query from each cluster (a total of 10K queries) and use these queries and associated documents as labels.
For the distractor, we randomly pick 3 queries from each cluster and get their documents. (documents from a total of 30K queries).
We exclude documents that have already been picked as positive for a distractor.

\subsubsection{Trainset}
Given the clustered result, we pick all queries from each cluster to obtain the trainset-small.
We construct the trainset by adding additional search logs into trainset-small.
For both sets, we exclude queries that are used in the positive and distractor of the testset and the validset.
Titles appeared in the positive and distractor of both testset, and the validset can be included.

\subsection{Details of Vocabulary Set Differences}
\label{a:vocab-diff-details}

This section describes the details and the effect of vocabulary set differences in our study.
Each record in the pre-training dataset contains at least 1 unigram of the corresponding unigram vocabulary set.
As mentioned, the two vocabularies have the same size of 100K, the ratio of intersection is 87.1\% (87,107 unigrams are shared). During the fine-tuning, the unigram vocabulary is not explicitly used as a unit of label.

We expect the effect of the difference in the vocabulary during the pre-training to be a non-significant impact, as many of the unigrams are shared between the two, since their vocabulary selection criteria are the "most frequent unigrams" from each vocabulary construction dataset of a natural language corpus.
This can introduce a larger number of associated sentences from the corpus for pre-training, as the probability of vocabulary occurrences is higher, and can lead to similar pre-training dataset records in terms of contained unigram vocabulary between the two vocabulary sets.

To that end, we expect the result of this study to be primarily differentiated by the finer-grained semantics of the pre-training corpus, as many lexical unigram vocabulary used for selecting data records are shared between the two, which can reduce the variations of selection results, as well as other pre-training configurations, instead of the difference in the vocabulary sets.

\subsection{Fine-tuning Configurations}
\label{a:ft-configs}

\subsubsection{Losses}
\label{a:losses}

We fine-tuned the EMLM model with a fine-tuning dataset for the ranking objective to obtain an ESPLADE model.

Unlike the original SPLADE paper, models were trained using only in-batch negative loss, which is formulated as:
\begin{equation}
\label{eq:1}
    \mathcal{L}_{\text{in-batch}} = - \frac{1}{|B|} \sum_{(q,d^+)} \log \frac{e^{s(q,d^+)}}{e^{s(q,d^+)} + \sum\limits_{d^- \in B} e^{s(q,d^-)}}
\end{equation}
where $s(q,d)$ represents the similarity score between query $q$ and document $d$, $d^+$ is the relevant document, $d^-$ are negative samples drawn from the batch $B$, and $|B|$ is the batch size. This decision was made to simplify the sampling process.
As a result, the Equation \ref{eq:1} is equivalent to the N-pair loss objective \citep{sohn2016improved}.

For regularization, the SPLADE model employs the FLOPS loss, which is defined as:
\begin{equation}
    \mathcal{L}_{\text{FLOPS}}(T) = \bar{w}^{(T)} \cdot \bar{w}^{(T)}
\end{equation}
where $\bar{w}^{(T)}$ denotes the mean representation of the output embeddings across all texts in $T$, and $\cdot$ is the inner product \cite{paria2020minimizing}. This loss applies separately to each of the Q, D representation batches.

In contrast, the ESPLADE model utilizes the joint FLOPS regularization loss, which is defined as:
\begin{equation}
    \mathcal{L}_{\text{jFLOPS}}(Q, D) = \bar{w}^{(Q)} \cdot \bar{w}^{(D)}
\end{equation}

This formulation ensures that the embeddings of queries and documents are effectively regularized, aligning their representations in the embedding space for improved retrieval efficiency.

\subsubsection{Hyperparameters}
\label{a:hps}

\begin{table*}
\centering
\small
  \begin{tabular}{l|ccccc}
    \toprule

model & val L0\_q & val L0\_d & val MRR @10 & val R@10 & val R@100\\
\midrule
overlap-repeat-lr-l & 10.22 & 29.7 & 0.9252 & 0.7234 & 0.8783\\
overlap-repeat-lr-h & 11.71 & 31.27 & 0.9309 & 0.7266 & 0.8759\\
indep-repeat-lr-l & 9.32 & 25.84 & 0.9289 & 0.7207 & 0.8723\\
indep-repeat-lr-h & 8.689 & 27.76 & 0.9304 & 0.7212 & 0.8716\\
indep-uniq-lr-l & 9.365 & 30.09 & 0.9282 & 0.7222 & 0.8743\\
indep-uniq-lr-h & 8.78 & 25.55 & 0.9312 & 0.721 & 0.8727\\

  \bottomrule
\end{tabular}

\caption{\textbf{Validset Result of Fine-tuned Models.} \small{
All models with 100K output vocab and a batch size of 3072.
The top-k masking options (we refer as q\_K and d\_K ) is from \cite{yang2021sparsifying} is set to 1000, 2000 respectively.
The joint FLOPS regularizer weight (we refer as L j) is from \cite{dudek2023learning} is set to 5.
The validset result is based on the best R@10 checkpoint.}}
\label{tab:validset-pref}

\end{table*}

We used the learning rate of 1e-4 with a linear learning rate scheduler. We use the warmup steps of 10,000.
For the optimizer, we used AdamW with weight decay of 0.01.
We used 16-bit (mixed) precision training.

ESPLADE, rand-vocab-100K model has 100K-sized output vocabulary, we trained with a 3072 batch size to 400K steps.

The validset results of fine-tuned models are presented in Table \ref{tab:validset-pref}.

\clearpage

\section{Backgrounds and Details for Q, D Pruning with MLM Logit Score}
\label{a:q-d-pruning}

The ranking loss and FLOPS regularization loss can be opposing forces in the training, each aiming to learn effective and efficient representations, respectively.
Since the two losses are combined in the finetuning, it can be difficult to obtain representations of maximal efficiency alone, as this can increase the ranking loss by decreasing the amount of information in the representations.

We applied static pruning methods for LSR \citep{lassance2023static} to obtain further retrieval efficiency. Specifically, we keep the top-k terms and drop other lower scored terms from the final learned sparse representations.
The MLM logit score is used.

In previous work \citep{buttcher2006document}, the target of static pruning is to remove terms in the inverted index, and the static pruning for LSR can follow a similar fashion \citep{lassance2023static}.
However, different from the traditional lexical sparse retrieval, the terms of LSR are learned, transform lexical representations, and the bi-encoder structure is trained to project Q and D vectors into the same embedding space. Hence, the gap between the Q and D term counts tends to be closer, compared to the lexical Q and D term counts~\footnote{Rough example can be found in Table~\ref{finetuning-pref-table} and Table \ref{tab:eval-set} show respectively.}.

Due to this, we applied static pruning for both documents as well as queries to further obtain retrieval efficiency, which is similar to the first matching setup discussed in \cite{lassance2024two}.
Throughout this paper, we use qk and dk, which show the max size of Q and D terms to represent each Q and D, which only includes the terms with the top-k highest MLM logit score.

\clearpage

\section{Statistics of EMLM Logit Vectors}
\label{a:statistics-emlm-word-embeddings}

\begin{table*}[ht]
\footnotesize
    \centering

  \setlength\tabcolsep{1.5pt}
  \begin{tabular}{lccccccccc}
    \toprule

& \multicolumn{3}{c}{doc-score-avg topk} & \multicolumn{3}{c}{doc-score-std topk} & \multirow{ 2}{1.5cm}{\centering {logit-score-std}} & \multirow{ 2}{1.5cm}{non-neg-terms-avg} & \multirow{ 2}{1.5cm}{non-neg-terms-std}\\
emlm model name & 10 & 100 & all & 10 & 100 & all &  &  & \\
\midrule
emlm-ptd-overlap-repeat-lr-l & 11.9609 & 8.4609 & -1.4414 & 2.66 & 1.63 & 2.54 & 5.82 & 42,791 & 36,406\\
emlm-ptd-indep-repeat-lr-l & 11.6172 & 8.2891 & -1.0781 & 2.65 & 1.56 & 2.37 & 4 & 40395 & 33198\\
emlm-ptd-indep-uniq-lr-l & 11.6719 & 8.2891 & -1.0332 & 2.67 & 1.58 & 2.35 & 4.33 & 41872 & 35066\\
\midrule
emlm-ptd-overlap-repeat-lr-h & 7.7852 & 4.4375 & -6.8008 & 2.25 & 1.53 & 3.02 & 6.77 & 14712 & 23409\\
emlm-ptd-indep-repeat-lr-h & 2.3359 & -1.3838 & -12.6328 & 2.86 & 1.73 & 2.94 & 9.49 & 9683 & 23754\\
emlm-ptd-indep-uniq-lr-h & 4.8438 & 1.084 & -10.25 & 2.88 & 1.75 & 2.94 & 9.65 & 17436 & 32712\\

  \bottomrule
\end{tabular}

\caption{\textbf{Stats of EMLM Models' logit vectors.} The logit vectors are from 164,740 WordPiece tokens from 8,000 positive documents from validset, where each logit vector has 100K dimensions (originally described in Section \ref{sec:emlm-logit-vector} of the main text).
}
\label{tbl:stats-emlm-models-by-lr}
\end{table*}

BERT word embeddings are contextualized representations of given words in a fixed dimension.
We used EMLM word embeddings, which are higher-dimensional logit vectors for word predictions, constructed based on BERT word embeddings.
In normal SPLADE, each logit index of the logit vector is mapped to one of the tokens supported by the tokenizer, as tokens are the atomic unit of predictions there.
This is similar in EMLM/ESPLADE, yet the specification of tokens is defined in a custom manner, such as a designated set of unigrams, which can be independent from the tokenizer.
The EMLM logit vectors will be pooled and used for ranking fine-tuning.
See Section \ref{sec:emlm-logit-vector} of the main text for the volume, source, and dimension of logit vectors.

Table~\ref{tbl:stats-emlm-models-by-lr} shows statistics of EMLM logit vectors. The description of types of statistics in the tables is:

\begin{itemize}
  \setlength\itemsep{0.5em}

\item {"non-neg-terms-avg" (Non-negative Terms Average)} is the average of the non-negative logit index count of each WordPiece token in documents;

\item {"non-neg-terms-std" (Non-negative Terms Std)} is the standard deviation of the non-negative logit index count of each WordPiece token in documents;

\item {"doc-score-avg" (Document-wise Score Average)} is the average of the average of each document’s top-k scores;

\item {"doc-score-std" (Document-wise Score Std)} is the average of the standard deviation of each document’s top-k scores;

\end{itemize}

\subsection{The Trend in Lower Learning Rate}

\textbf{The Repetition of Pre-training Corpus (emlm-ptd-indep-repeat-lr-l vs. emlm-ptd-indep-uniq-lr-l).}
In Table~\ref{tbl:stats-emlm-models-by-lr}, the doc-score-avg and doc-score-std of both emlm-ptd-indep-repeat-lr-l and emlm-ptd-indep-uniq-lr-l model is similar, while the emlm-ptd-indep-uniq-lr-l has slightly higher logit-score-std, non-neg-terms-avg, and non-neg-terms-std.
The definition of metrics allow us to understand that MLM predictions on the repeated dataset have a slightly smaller number of score diversity on the individual logits, and slightly less number and less variance of non-negative terms\footnote{Currently, we expect this is due to overfitting, as repetition in the corpus can make predictions easier, the model can have a higher margin to positive logit and other logits on the basis of classification loss, with multiple steps of optimization in a similar input context. This can decrease entropy in score values and entropy of non-negative scored logit indexes of the logit vector, as the model provides predictions that are too confident and less adaptable. \citet{szegedy2016rethinking} discussed cross-entropy loss and overfitting when describing label-smoothing regularization.}.

\textbf{The Contents of Pre-training Corpus (emlm-ptd-overlap-repeat-lr-l vs. emlm-ptd-indep-repeat-lr-l).}
For the case of emlm-ptd-overlap-repeat-lr-l, this model uses a different pre-training corpus from emlm-ptd-indep-*, we can see its logit-score-std, non-neg-terms-avg, and non-neg-terms-std are higher.

This dataset has repeated, yet does not show smaller values on these metrics, as shown in the case of emlm-ptd-indep-repeat-lr-l in emlm-ptd-indep-*, which can represent that the contents of the pre-training dataset also affect values on these metrics, obviously.
\footnote{The pre-training dataset of emlm-ptd-overlap-repeat-* is overlapped with trainset, which has a similar distribution to the validset. One possible interpretation of metrics is that the MLM model gives higher logit-score-std, non-neg-terms-avg, and non-neg-terms-std when the input data is in-fine-tuning.
For a speculation, we can assume that masking the same data with different masking patterns can affect such activations of logits, as it sets different labels for similar input contexts, which can make the model more confident and less adaptable to the input of in-fine-tuning data.
Note that in the field of machine learning, label-smoothing regularization~\citep{szegedy2016rethinking} is introduced to train models for less confident and more adaptable.
}.

\subsection{The Trend in Higher Learning Rate}

The models with *-lr-h in the Table~\ref{tbl:stats-emlm-models-by-lr} are trained with a higher learning rate.
To compare the emlm-ptd-indep-uniq-lr-h and emlm-ptd-indep-repeat-lr-h models, the similar trend observed as of *-lr-l, logit-score-std, non-neg-terms-avg, and non-neg-terms-std is decreased.
However, the logit-score-std of emlm-ptd-overlap-repeat-lr-h is lower than aforementioned two models, which is a different trend from *-lr-l.
The result shows that increasing the learning rate increases the logit-score-std metric, while such an increase is larger when the corpus is general~\footnote{
This can be interpretable as the higher learning rate promotes higher variance on the use of logit indices in terms of scores, and even more when the pre-training corpus is general.

For all models, compared to the models with a lower learning rate, logit-score-std is increased; however, non-neg-terms-avg and non-neg-terms-std are decreased.
A possible connection to SPLADE fine-tuning is that a higher learning rate can foster sparsity of high-scored logits by having a large margin to other logits, as higher logit-score-std shows, yet can have less entropy in their logit vectors, as decreased non-neg-terms-avg, non-neg-terms-std shows.
These properties can have a connection to the sparsity regularization loss and contrastive loss of SPLADE fine-tuning.
The main experimentation result of this paper can be discussed in this direction.
}.

The doc-score-avg score has a difference, orderable by emlm-ptd-indep-repeat-lr-h < emlm-ptd-indep-uniq-lr-h < emlm-ptd-overlap-repeat-lr-h. While lower compared to the *-lr-l.
This indicates training with a higher learning rate for classification yields a model with lower logit scores on average.
While comparably higher variances of scores (logit-score-std), and allowing the model to form a decision boundary using a smaller number of non-negative terms.
Overall, this can be affected by cross-entropy loss as it promotes the top-1 answer and makes a margin to others.

\subsection{Comprehensions for Pre-training Data and doc-score-avg Metric}

In Table~\ref{tbl:stats-emlm-models-by-lr}, the higher doc-score-avg of emlm-ptd-overlap-repeat-lr-h is observed~\footnote{
This can be related to the result of both in-fine-tuning pretraining and in-fine-tuning validation data, where such alignment can increase the absolute scale of scores.
}.
The model emlm-ptd-indep-repeat-lr-h shows the lowest scores on doc-score-avg metric~\footnote{This can be related to the combination of a higher learning rate and repeated and general pre-training data.}.
Model with non-repeated general pretraining data emlm-ptd-indep-uniq-lr-h shows a medium level score scale for doc-score-avg metric~\footnote{
This might represent that diverse prediction labels from non-repeated pre-training data are helpful for increasing the score scale, even in the data is general.
}.

\clearpage

\section{The Effect of Pre-training Steps}
\label{apdx:pretraining-steps-effect}

\input{asset-8-2-emlm-pretraining-metrics-longer-steps.tex}

\begin{table*}[ht]
\centering

\setlength{\tabcolsep}{3.75pt}

\scriptsize

  \begin{tabular}{l|cccc}
    \toprule

emlm model & emlm-ptd-indep-uniq-lr-h-600K & emlm-ptd-indep-uniq-lr-h-1.2M & emlm-ptd-indep-uniq-lr-h-2.9M\\
\midrule
non-neg-terms-avg & 17,436 & 18,925 & 26,976\\
\midrule
non-neg-terms-std & 32,712 & 37,624 & 43,582\\
\midrule
doc-score-avg &\\
topk=10 & 4.8438 & -13.5547 & -18.9375\\
topk=100 & 1.084 & -17.1406 & -22.1719\\
topk=all & -10.25 & -28.1406 & -31.5938\\
\midrule
doc-score-std &\\
topk=10 & 2.88 & 2.8 & 2.6\\
topk=100 & 1.75 & 1.68 & 1.52\\
topk=all & 2.94 & 2.92 & 2.48\\
\midrule
logit-score-std & 9.65 & 26.77 & 59.03\\
  \bottomrule
\end{tabular}
  \caption{\textbf{Stats of EMLM Logit Vectors on Different Pre-training Steps.}}
  \label{tab:mlm-logit-vec-training-steps-diff}

\end{table*}

Losses \& accuracies of EMLM pre-training on longer steps are depicted in Figure~\ref{fig:emlm-pretraining-losses-longer-steps}.
Table \ref{tab:mlm-logit-vec-training-steps-diff} shows the statistics of logit vectors from EMLM model checkpoints at the steps of 600K, 1.2M, 2.9M. The total training step configured for training is 2.9M.
The learning rate decay method is linear.
We employed nearly 3 billion web titles for this pre-training, which are all unique.
Hence, during pre-training, each model sees approximately 614M, 1.2B, and 2.9B training records, as we used a 1024 batch size for each training step.
The fine-tuning of emlm-ptd-indep-uniq-lr-h-1.2M and emlm-ptd-indep-uniq-lr-h-2.9M failed after drastic shrinking of queries and documents' activated terms.

\clearpage

\section{Analysis of Fine-tuned Representations}
\label{a:analysis-on-pretraining-effect}

\begin{table*}[ht]
\centering
\setlength{\tabcolsep}{3pt}
\footnotesize
  \begin{tabular}{lcccc|cccc|cccc}
    \toprule

 & \multicolumn{4}{c}{term\_occ\_threshold=10000} & \multicolumn{4}{c}{term\_occ\_threshold=5000} & \multicolumn{4}{c}{term\_occ\_threshold=1000} \\
 & dk=0 & dk=20 & dk=10 & dk=5 & dk=0 & dk=20 & dk=10 & dk=5 & dk=0 & dk=20 & dk=10 & dk=5\\
\midrule
rand-vocab-100K & \\
logit-score-std & 0.5567 & 0.4827 & 0.34 & 0.2622 & 0.5428 & 0.4897 & 0.3468 & 0.2567 & 0.3714 & 0.4775 & 0.3518 & 0.2505\\
logit-cnt & 24893 & 15746 & 5864 & 953 & 33150 & 25964 & 17089 & 7077 & 60516 & 34206 & 29136 & 23283\\
\midrule
\multicolumn{5}{l}{overlap-repeat-lr-l} & \\
logit-score-std & 0.4278 & 0.5193 & 0.3955 & 0.3028 & 0.3864 & 0.5488 & 0.4156 & 0.2986 & 0.3567 & 0.5001 & 0.4433 & 0.3016\\
logit-cnt & 42290 & 12960 & 5170 & 684 & 74511 & 25585 & 14925 & 5941 & 98909 & 46618 & 33803 & 26620\\
\midrule
\multicolumn{5}{l}{overlap-repeat-lr-h} & \\
logit-score-std & 0.439 & 0.4798 & 0.356 & 0.2714 & 0.4123 & 0.5069 & 0.368 & 0.2696 & 0.3927 & 0.4926 & 0.3943 & 0.268\\
logit-cnt & 46549 & 12126 & 4154 & 500 & 79491 & 27266 & 14315 & 5132 & 99446 & 52926 & 39084 & 28433\\
\midrule
\multicolumn{5}{l}{indep-repeat-lr-l} & \\
logit-score-std & 0.3833 & 0.5373 & 0.4017 & 0.2996 & 0.3699 & 0.5656 & 0.4168 & 0.2927 & 0.3629 & 0.5328 & 0.443 & 0.291\\
logit-cnt & 51696 & 13016 & 5186 & 722 & 81757 & 25401 & 15314 & 5923 & 96891 & 41474 & 32817 & 26708\\
\midrule
\multicolumn{5}{l}{indep-repeat-lr-h} & \\
logit-score-std & 0.491 & 0.5273 & 0.3973 & 0.3 & 0.4539 & 0.5516 & 0.413 & 0.2956 & 0.4031 & 0.4951 & 0.4333 & 0.2939\\
logit-cnt & 33849 & 12749 & 5110 & 758 & 57998 & 25299 & 14865 & 5920 & 89319 & 46821 & 33137 & 26416\\
\midrule
\multicolumn{5}{l}{indep-uniq-lr-l} & \\
logit-score-std & 0.3725 & 0.5302 & 0.3972 & 0.2999 & 0.3643 & 0.5586 & 0.4128 & 0.2925 & 0.3608 & 0.521 & 0.437 & 0.29\\
logit-cnt & 54133 & 13068 & 5097 & 700 & 83303 & 25535 & 15337 & 5855 & 97053 & 42804 & 32974 & 26820\\
\midrule
\multicolumn{5}{l}{indep-uniq-lr-h} & \\
logit-score-std & 0.5549 & 0.5466 & 0.4057 & 0.3007 & 0.5025 & 0.5769 & 0.4238 & 0.297 & 0.4141 & 0.5622 & 0.4498 & 0.2963\\
logit-cnt & 27733 & 13070 & 5107 & 749 & 49144 & 25420 & 15078 & 5891 & 87467 & 39131 & 32977 & 26625\\

  \bottomrule
\end{tabular}

  \caption{\textbf{Stats for Average of the
Standard Deviation of Individual Logit Indices’ Scores Appearing in All Evaluation Set Documents.}
  \small{
  \vspace{2px}\\ *As depicted in the Table \ref{tab:eval-set}, the number of documents is 20,372,952. \vspace{2px}\\ *logit-score-std is the average of the standard deviation of individual logit indices’ scores appearing in all Evaluation Set documents. \vspace{2px}\\ *term\_occ\_threshold is the minimum logit (term) occurrence threshold for the stats. Only using scores of logits that appear the same or greater than this value is used to calculate the std values, hence affecting to logit-score-std. \vspace{2px}\\ *logit-cnt is the number of logits that met the term\_occ\_threshold conditions. \vspace{2px}\\ *Note that the purpose of multiple term\_occ\_threshold conditions and logit-cnt values is to check the statistical significance of the observation. \vspace{2px}\\ *rand-vocab-100K model was fine-tuned from the random base model. The random base model has the same physical architecture as the EMLM model, however, without EMLM pre-training. Prior BERT pre-training was applied. The output MLM FC layer was initialized randomly.
  }
  }

  \label{tab:eval-set-logit-score-std}
\end{table*}

\subsection{Logit(Term)-Wise Statistics}

\begin{sloppypar}
Table~\ref{tab:eval-set-logit-score-std} shows stats for activated terms on the evaluation set.
In the conditions of term\_occ\_threshold=10000 and term\_occ\_threshold=5000, in the unpruned setting(dk=0), the random vocab model, rand-vocab-100K shows higher logit-score-std scores compared to other models.
This can imply, as random suggests, higher randomness in the score of their activated terms. %
In the pruned setting~(dk=20, dk=10, dk=5), the trend generally flipped,
ESPLADE models show generally higher logit-score-std scores.
This can be interpretable as compared to its random base model (that is used for pre-training the random vocab model), the pre-trained EMLM model trained with the MLM classification task gives larger variance in the scores of top-k logits. %
\end{sloppypar}

In the conditions of term\_occ\_threshold=10000 and term\_occ\_threshold=5000, comparing the ESPLADE models, in dk=0, increasing the learning rate increases the ESPLADE logit-score-std scores, and these scores show some correlations to the EMLM logit-score-std depicted in the Figure~\ref{finetuning-pref-figure} in terms of relative scale.

In the conditions of term\_occ\_threshold=10000 and term\_occ\_threshold=5000, in the pruned setting~(dk=20, dk=10, dk=5), indep-* models, increasing the learning rate does not much change the ESPLADE logit-score-std scores.
Under the same conditions and pruning settings, in the case of overlap-repeat-*, increasing the learning rate decreases the logit-score-std scores.
Under the same conditions and pruning settings, the logit-score-std score of the overlap-repeat-lr-h model is especially low compared to other models, and the retrieval effectiveness of this model becomes low in a large margin in most strict pruned settings, as depicted in Figure~\ref{finetuning-pref-figure}.

\subsection{Query and Document-Wise Statistics}
\begin{table*}
\centering
\setlength{\tabcolsep}{3.75pt}
\footnotesize
  \begin{tabular}{l|ccc|cccc|cccc}
    \toprule

model & \multicolumn{3}{c}{L0\_q-std} & \multicolumn{4}{c}{L0\_d-std} & \multicolumn{4}{c}{pos-ls-len-std} \\
 & full & top7 & top5 & full & top20 & top10 & top5 & full & top20 & top10 & top5\\
 \midrule
rand-vocab-100K & 9.88 & - & 0.22 & 29.43 & 2.51 & 0.35 & - & 11852 & 7603 & 4627 & -\\
overlap-repeat-lr-l & 9.52 & 0.95 & 0.35 & 56.87 & 2.36 & 0.39 & - & 10599 & 6779 & 4055 & -\\
overlap-repeat-lr-h & 10.33 & 0.72 & 0.28 & 57.09 & 1.92 & 0.3 & 0.05 & 10402 & 6307 & 3757 & 2395\\
indep-repeat-lr-l & 7.94 & 1.06 & 0.44 & 115.35 & 3.08 & 0.51 & 0.08 & 10242 & 6909 & 4152 & 2669\\
indep-repeat-lr-h & 8.05 & 1.19 & 0.52 & 57.45 & 2.75 & 0.48 & 0.08 & 10376 & 6812 & 4149 & 2611\\
indep-uniq-lr-l & 7.68 & 1.03 & 0.4 & 118.05 & 2.94 & 0.49 & 0.07 & 10448 & 6887 & 4139 & 2675\\
indep-uniq-lr-h & 7.11 & 1.19 & 0.52 & 52.39 & 3.31 & 0.57 & 0.09 & 10026 & 7001 & 4206 & 2647\\

  \bottomrule
\end{tabular}
  \caption{\textbf{Stats for Activated Terms on the Evaluation Set.}
  \small{
  \\ *The number of queries is 8,936 and the number of documents is 20,372,952 (Table \ref{tab:eval-set}). \vspace{2px}\\ *qk, dk is the Q, D top-k pruning option. \vspace{2px}\\ *L0\_q-std, L0\_d-std are the standard deviation from the term count of individual Q, D. \vspace{2px}\\ *pos-ls-len-std is the standard deviation of postings list length (i.e., the standard deviation of logit index-wise occurred terms count of D).
  }
  }
  \label{tab:eval-set-act-term-stats}
\end{table*}

Table~\ref{tab:eval-set-act-term-stats} shows stats for activated terms on the evaluation set.
The table shows L0\_q-std, L0\_d-std, which are the standard deviation from the term count of individual Q, D.
Among the ESPLADE models with the lr-h settings, the overlap-repeat-lr-h model generally shows larger or similar L0\_q-std, L0\_d-std values compared to the indep-repeat-lr-h and indep-uniq-lr-h models, in an unpruned setting. However, this trend becomes the opposite when the pruning option is applied.

The pos-ls-len-std shows the standard deviation of postings list length.
Among the ESPLADE models with the lr-h settings, in an unpruned setting, the score is relatively similar between models; on the contrary, in a pruned setting, the overlap-repeat-lr-h shows a lower score.

In the case of rand-vocab-100K, when unpruned, the L0\_q-std is similar to other models, but the L0\_d-std is lower by a large margin compared to other models. We expect this as random initialization promotes overall flattened activations of logits and scores.
\footnote{
\citet{kim2025role} shows that the effect of such distributions results in low retrieval effectiveness when pruned.
}
The pos-ls-len-std metrics of the rand-vocab-100K model are higher in both unpruned and measured pruned ranges. %

\clearpage

\section{Postings List Length Variances and Retrieval Effectiveness, Efficiency}
\label{a:pos-ls-len-var-and-retrieval-ee}

Section~\ref{pos-ls-len-dist-and-retrieval-ee} shows that the relationship of higher variances of postings list length to higher retrieval effectiveness and decreased retrieval efficiency.
This can be possibly related to the inherent lexical non-uniformity of natural language sentences and the following semantic non-uniformity of learned sparse representations derived from such natural language sentences.
Such properties of a natural language corpus can inevitably introduce a postings list of \textit{short-head latent terms} that have long lengths due to their commonality, consequently have a high std value of postings list length.
Conceptually, this can contrast with a short length of the postings list made by comparably \textit{long-tail latent terms}.
The \textit{short-head latent terms}, which occur frequently, can lead to a negative effect on retrieval efficiency, resulting in higher FLOPS, as traversal on the longer postings list occurs more frequently.

We expect the non-uniform nature of lexical representations can be one cause for this issue, and on the other hand, the relevance labels of ad-hoc IR can have latent and general patterns of mapping Q and D; and we expect the existence of such a pattern can also contribute to this kind of non-uniformity. As training would be guided by such a pattern, and that means the guide terms that are used by such a general pattern would have frequent activations. For instance, the current Expanded-SPLADE loss architecture might assign arbitrary (less semantically meaningful) terms to bridge the concept of Q and D while accepting the FLOPS regularization, as discussed by~\citet{kim2025role}.
Overall, such non-uniformity originated from the representations and the learning, can be a \textit{Zipfian-like property} in the Expanded-SPLADE postings list\footnote{
Zipfian-like semantic structure of a general natural language corpus, where literatures characterize them as "Small World Structures" whose structures have strong local clustering and low distances, having hubs that have a high number of connections~(e.g.,~\citet{steyvers2005large, cancho2001small}).
Although our learned sparse representations are trained for ranking in bi-encoders form, such semantic representational characteristics seem involved and maintained, as the input is a natural language corpus.
And broadly, this concept seems to be connected to the notion of non-speciality and speciality words of lexical representations in models of indexing, like the 2-Poisson model of indexing~\citep{harter1975probabilistic1, harter1975probabilistic2}.
}.

\clearpage

\section{Understanding the Limits of EMLM Pre-training for IR Fine-tuning}

The discrepancy in pre-training can occur since pre-training can be viewed as a form of regularization \citep{erhan2010does}.
To follow this view, regularization reduces the representational capacity of the model for generalization in the fine-tuning. However, too much regularization reduces the variance of the model could result in underfitting at fine-tuning.
Thus, in this case, we need to measure the model's ability to represent literally or semantically similar representations well, with different emphases of logits index and score, for high variance in high-dimensional output, to maintain representational capacity for the retrieval fine-tuning, and avoid underfitting at fine-tuning by strong regularization at pre-training.
The reason is, as discussed previously, the contrastive fine-tuning of the IR task needs to construct the mapping of query representation and document representation in the designated embedding space, while the MLM pre-training primarily focuses on the learning of query or document representation itself~\footnote{
For example, the representation of "where is the capital of Canada?" should be similar to "where is the capital of Japan?", and can be less similar to "Ottawa" in the MLM pre-training, as the typical MLM does not map Q and D together, only learns the overall context of individual sentences. Some similarity can also be provided between "where is the capital of Canada?" and "Ottawa" as both inputs can be co-occurring and can be contained in the input context. However, at the same time, the first case still exists.
}~
\footnote{
This understanding led us to think about a method to evaluate whether the model is underfitting for the retrieval downstream task (i.e., by too much regularization from too much MLM pre-training) as well as evaluate whether the model will be overfit in the retrieval downstream task (i.e., by too little regularization from too little MLM pre-training).
For instance, we can think of using a clustering-based approach, by measuring how well clustering results based on high-dimensional vectors, created by applying pooling to EMLM logits, can reconstruct ground-truth clustering results using a general-purpose embedding model like SBERT \citep{reimers2019sentence}.

If we assume that a higher inner-cluster variance implies the presence of more additional latent terms available for fine-tuning (i.e., increased degrees of freedom or variances that can be used for fitting the downstream retrieval task), then when evaluating EMLM based on cluster recoverability, we could also measure the inner-cluster variance.
If two EMLMs show similar clustering recoverability accuracy, we could investigate whether the one with higher inner-cluster variance can be preferable for fine-tuning.

That model's representations might have a characteristic that optimally balances between maximizing clustering recoverability accuracy (maximizing expressiveness) and maximizing intra-cluster variance, which means that even when semantically similar sentences are represented, the model can assign high scores to different logits, rewarding more for diverse representations that are usable for the fine-tuning ranking loss objectives, considering having higher representational capacity for the fine-tuning.
As future work, we could test further experimentation to check whether this measurement approach can be accepted.
}
.

\clearpage

\section{Future Work and Directions}

When training SPLADE models, the discriminative forces and the sparsification forces in the loss formula can be opposing objectives, which can make foundational trade-offs in current MLM-based pre-training in LSR.
This can hinder the application of advanced language representation methods while minimizing the loss, such as in the case of dense retrieval, where dense embeddings are the vanilla output of transformer-based models; however, high-dimensional representations for sparse embeddings are typically not.

In this regard, we expect one of the major bottlenecks of learning sparse representations for ranking to be the transformation of a dense vector to an MLM vector, where classificational output prioritizes the top elements while relatively ignores other~\footnote{
This can be viewed as an additional bottleneck layer of the model, in addition to the interactional bottleneck of Q, D representations compared to the cross-encoder models~\citep{nogueira2019passage, luan2021sparse}.
}.
This may be part of the source of the efficiency gain of LSR compared to dense retrieval.

Aiming to further improve the LSR, we encourage the community to explore diverse ways to implement it.
Here, we share some of the prospective topics to complement expressiveness and sparsity to improve the effectiveness and efficiency of the current MLM-based LSR.

\subsection{Training with Larger Output Vocabulary on Expanded-SPLADE Models}

The Zipfian-like distribution can be problematic in terms of efficiency in the IR task.
In the IR task, we collate documents into the collection and convert them to the inverted index; hence, distributions of multiple documents are comprised, resulting distribution has more higher variance than that of the original individual records.

In the learned sparse retrieval, the Zipfian-like distribution of language causes a bottleneck for retrieval efficiency.
Serving on the inverted indexes is efficient compared to the Nearest Neighbor Search, but its efficiency is naturally related to the term distribution of the inverted indexes revealed by the Document Frequency (DF) or postings list length of the term.

Alleviation of this issue can inovolves many approaches, can include expanding the output dimension to have more fine-grained terms to represent common concepts in the language as discussed in the~\citet{kim2025role}\footnote{For this purpose, we can think of the role of sparse neural matching as a decomposer, or technically, a hash function that divides common concepts into fine-grained concepts.}.
Other approaches can include two-step retrieval, first-step retrieval using a pruned index, and second-step accurate scoring using unpruned vectors to the result of first-step retrieval~\citep{lassance2024two, bruch2024efficient}.

The one possible approach, use of larger vocabularies, as recent research of \citet{kim2025role} discussed that the size of vocabularies has positive correlation to the models performance when retrieval using Q, D pruning, and they noted that larger vocabularies can lead to the larger training GPU memory consumption of storing larger MLM logit vectors per each input tokens prior to the max pooling operations. Reducing the size of intermediate MLM logit vectors can be important to alleviate this issue. And for that, we can consider dense embedding level pooling strategies, such as mean pooling for all of the output dense vectors. Mean pooling is a widely used dense embedding level pooling strategy used in models like Sentence-BERT~\citep{reimers2019sentence}. This can reduce the effectiveness of the models as contextualized embeddings are averaged; however, we expect there are some benefits of using larger output vocabularies instead, which is directly related to the informativeness of output sparse representations~\citep{kim2025role}.
In addition, such pre-pooling, an MLM logit vector level pooling, can reduce the memory and computational cost of serving the model.
Word vector elimination methods~(e.g., \citet{goyal2020power}) can also be applicable for the same motivations.

\subsection{Pre-Training with Instance-Agnostic and Inter-Class Discrimination}

Less reliant on a self-supervised pre-training approach, and using normal supervised pre-training can be considered.
Throughout the paper, we can see that the MLM task is not very aligned to the ranking fine-tuning as it is inherently an intra-instance discrimination task in the form of self-supervised learning, instead of an inter-instance discrimination task that involves information beyond a single target instance.

Typical supervised learning tasks involve an inter-class discrimination task, which aims to predict labels; even predicting labels is only a scaffold for obtaining latent representations, such as dense embeddings.
For instance~\citet{el2021training} used CLS token embeddings of Vision Transformers, which are used for the class predictions of an image~\citep{dosovitskiy2020image} for image retrieval by fine-tuning on contrastive loss.
 
In an industrial setting where there is abundant data, search queries and clicked documents can be used for label signals for conducting normal supervised pre-training (not self-supervised pre-training) of language models for IR tasks.

On the other hand, \citet{an2023unicom} shows different approaches for constructing such labels; they clustered a pre-training dataset to make 1 million pseudo classes based on features obtained from foundation models and used as a basis for the image retrieval training, which can be an alternative approach where large-scale label-based supervision is difficult to attain.
Note that empirical results~\citep{radford2021learning, an2023unicom} show that, however, the quality of the label and hence the quality of the supervision signals are crucial; the training dataset volume and model performance tend to have a positive correlation, but the conversion ratio is not strong if the pre-training dataset is constructed with weak supervision.
This implies that an improvement of label quality is crucial, as well as increasing the volume of the pre-training dataset.

Fine-tuning of such high-dimensional embeddings from normal supervised pre-training can provide sparse embeddings aware of instance-agnostic, inter-class discrimination, instead of intra-instance discrimination, which might provide improved readiness for retrieval fine-tuning, compared to low readiness of MLM pre-trained BERT as pointed by~\citet{gao2021condenser}.

\subsection{The Boolean Model with Static Term Overlap Threshold and SPLADE}

In a search engine, the number of first matched document usually impact computational and memory cost, as heavier processing is needed for the matched document for ranking to calculate the relevance.
SPLADE can use the dot product for its similarity measure; however, training models to have a higher dot product similarity score does not explicitly align with such efficiency characteristics of search engines that are sensitive to the number of first matched documents.
For instance, in SPLADE, allowing the retrieval engine that first match any item that contains a specific term can introduce a larger number of first matched results as the vocabulary of SPLADE is a fixed dimension and is dense~\citep{kim2025role} compared to the vocabularies of natural language.
However, it is assumed that generally, the higher probability of a higher matching score of SPLADE Q, D vector is expected, if the number of intersecting terms increases, as it can contribute to the higher doc product score.

Such Q, D term co-occurrence allows us to use the concept of the Boolean Model~\citep{lancaster1973information} with term static overlap threshold criteria for further enhancing retrieval efficiency.
The Boolean Model is a classical IR model that accepts query that uses Boolean logic, and the retrieval is done by identifying satisfied documents of the given logic.
The Boolean Model with static term overlap threshold means, before the first matching, for each document with a given query, calculate the ratio of query terms that are matched from the document terms, and use the ratio and fixed threshold to determine the acceptance of a document for the first matching.
While the Boolean model is extended to support partial matching and term weights~\citep{salton1983extended}, LSR delegates such ability to parametric neural models such as BERT~\citep{devlin2019bert}, thus we can attempt to use the basic Boolean Model with the static term overlap threshold to reduce the number of first matched documents.
In this way, the number of first matched documents can be reduced.
And according to the aforementioned properties of SPLADE's sparse representations, we expect that much of the discarded document has a comparably low dot product similarity score.
\citet{won2025efficiency} shows an efficiency gain of this approach.

One step further, we can think of training sparse representations aligned for the retrieval that uses a Boolean Model with a static term overlap threshold.
An ideal model for this setting can represent positive Q and D with multiple co-occurring relatively medium-scored terms.
Conversely, depending on a small number of co-occurring relatively high-scored terms for representing positive Q and D cannot be aligned to the Boolean model with the static term overlap threshold.
Without a static term overlap threshold, the Boolean Retrieval that accepts documents with OR conditions for all query terms, which is equivalent to dot-product similarity-based retrieval, can result the comparably excessive first matching volume, less efficient by making the first stage ranker required to process a larger volume of document that has, for example, one of the single terms in queries.
In this case, applying the Boolean model with the static term overlap threshold for the aim to improve efficiency can result trade-off in retrieval effectiveness to some degree.

To make a model to represent positive Q and D with multiple co-occurring relatively medium-scored terms, friendly for the retrieval with a Boolean model with the static term overlap threshold, we can penalize part of the model training.
For instance,
j FLOPS Regularization loss~\cite{dudek2023learning} can be related to this purpose, as it regularizes the intersection of Q, D representations can balance intersectional co-occurrences of terms as well as intersectional sparsity.
In terms of enforcing the model to have discriminative power using primarily the logit index, instead of the logit score, we can use cosine similarity as a similarity measure instead of or in addition to the dot product during the training.
Such a model can lose comparably smaller discriminative power after applying the Boolean Model with a static term overlap threshold\footnote{However, there is a possibility that the models tend to output more terms in this case to maintain informativeness of representations where cosine similarity only concerns angle, as different scales of scores with the same angle are indistinguishable. Whether it is big or small, such a trade-off can exist as a similarity component in the loss changes.}.

Alternatively, we can think of modifying the FLOPS regularization loss formula to support term co-occurrence of positive Q and D along with term diversity, instead of only concerning with reducing the overall FLOPS score on the regularization loss term. For example, metrics like Jaccard distance or Hamming distance can be usable for formulating the loss value in order to direct models in such a direction.

\subsection{Configuring Embedding Spaces for Robust Parametric Bi-Encoders based Retrieval}

\subsubsection{Parametric IR Models and It's Limitations}

Parametric IR models can be prone to generalization of unseen queries and documents. %
Concretely, some unseen queries and documents can be problematic if they are difficult to be generalized from the training dataset of Parametric IR models which is comprised of seen queries and documents.
Non-Parametric IR models such as BM25~\cite{robertson2009probabilistic} can be less impacted by such an issue; however, they can be less flexible with respect to the matching performance from divergence of vocabularies, and inherent asymmetricity and semantics gap between queries and documents.

In essence,
when looking at this contrast of Parametric and Non-Parametric IR models,
one way to interpret such a contrast is,
the learning of prior knowledge other than the lexical representations is inherently latent, and the learning of a full spectrum is impractical due to the myriad combinational cases, and such learning is time sensitive and time decay, as the world it reflects is continuously changing.
This can be one of the characteristics of a generic, high-level view of the IR in the real world\footnote{e.g., Google Revisits 15\% Unseen Queries Statistic In Context Of AI Search. Search Engine Journal. 2025.
}.
This can mean there is a highly likely area of input space of queries and documents that is unseen and is difficult to generalize for the Parametric IR model, such as SPLADE.
This can lead to having false positivity or false relevancy between not related queries and documents.

To minimize such an issue, performing metric learning, notably contrastive learning properly to establish a robust embedding space can be important.
The contrastive learning produces an embedding, where the concept of contrastive learning discriminates one from the others. Such embeddings can be usable as applications like clustering and classification based on the nearest neighbour, and are also aligned to the retrieval.

The concept of learning \textit{metric} and learning \textit{contrast} seems inherently related to the IR, as the concept of a document's relevance to the query can be interpretable as a concept of metric or distance, and we make feature to make useful contrast between documents with respect to queries, where the degree of the contrastiveness can be a score of an irrelevancy.

The contrastive learning has been shown to be effective in LSR, yet rather than focusing on learning with robust pairwise positive relationships,
what approaches can we take in order to be less suffered by the return of the irrelevant result of false positive documents?

Consider an example that, in contrastive learning in IR, the two driving forces are attractive forces between a pair of Q and positive D, and repulsive forces between a pair of Q and negative D.
And these two forces can configure the structures of embedding spaces.
On the other hand, we currently have a hypothesis that Q and D, which are difficult to generalize, are more likely to produce an embedding that refers to the area of embedding space implicitly constructed by repulsive forces,
where the area of embedding space constructed by attractive forces can mainly focus on covering instances of Q and D generalizable from its training set.
We expect the above hypothesis can be related to a spurious correlation caused by less generalizable input. In a practical case, we observed that misspelled queries are matched to rare document titles.

Moreover, even if the patterns are in the dataset, the learnability can still be incomplete for some cases.
As pointed out by \citet{oh2017deep}, the N-pair metric learning loss objective is not always aware global structure of the embedding
space, as noted by \citet{kim2025role} in the context of LSR.
This can potentially hinder learning the whole relationship embodied in the dataset.

To summarize, the bottleneck of generalization in Parametric IR models is two-fold in terms of task coverage and learning methods.
One bottleneck is related to the inherent limitation of the training dataset coverage of the task.
And another bottleneck is, limitation of contrastive learning methods like N-pair metric learning, resulting in an incomplete utilization of latent knowledge embodied in the target dataset.

\subsubsection{Some Approaches for Robust Parametric Bi-Encoders}

To reduce the aforementioned false positive cases, training models to make query representations that are less interfere with other document representations, hence getting a lower similarity score seems important, for the input of unseen cases in the training dataset and the input difficult to generalize.

One possible approach for tackling that is making the positive instances more cohesive, thus having a larger margin with non-positive cases.%
\citet{yang2019improving} suggests ways to learning multilingual embeddings with Additive Margin Softmax for translation retrieval tasks, where Q and D expressions are semantically similar but different by their language.
They reported an improved performance in their task based on dense embedding representations.
And this can be conceptually similar to circumstances for an ad-hoc retrieval task, where Q, D are associated in terms of retrieval but mostly different in their lexical representations.
The concept of Additive Margin Softmax can be applicable for EMLM pre-training. However, we assume the effect for retrieval fine-tuning can be uncertain. Due to the observed pre-training and fine-tuning discrepancy.
Conceptually, the concept of margin can be applicable for contrastive fine-tuning as well; for instance, we can add a constant margin to the similarity score of a positive pair in the contrastive loss term.
This can enforce a larger dissimilarity to other groups by increasing the margin between other groups.

In addition to the above, we can employ a MoCo~\citep{he2020momentum} approach for leveraging large-scale negatives, similarly applied to the dense retrieval model like Contriever~\citep{izacard2021unsupervised}.
Using a larger scope of negatives can help to make representations more reliable.
ANCE~\citep{xiong2020approximate} is an approach for using a retriever as part of the training process for mining hard negatives.
However, mining hard negatives of full coverage is difficult, because the mining is bounded in the current dataset pool, thus not exhaustively included for unseen and hard-to-generalize inputs at test-time.

\clearpage

\end{document}